\documentclass[12pt]{article}
\usepackage{graphicx}
\usepackage{amsmath,amssymb}
\usepackage{hyperref}
\begin{document}
\def\be{\begin{equation}}
\def\ee{\end{equation}}
\def\ba{\begin{eqnarray}}
\def\ea{\end{eqnarray}}	
\def\l{\left}
\def\r{\right}
\def\fr{\frac}
\def\la{\label}
\def\d{\partial}
\def\vphi{\varphi}
\def\pr{\prime}
\def\mpl{M_{\rm p}}
\def\hub{\mathcal{H}}
\def\mbf{\mathbf}
\def\baselinestretch{1.2}


\author{Peter Gilmartin\thanks{pgilmartin01@manhattan.edu}, Bart Horn\thanks{bhorn01@manhattan.edu}}
\title{\bf\Large Observable relics of the simple harmonic universe}
\date{\normalsize\textit{Manhattan College, Department of Physics,\\ New York, NY 10471, USA}}

{\let\newpage\relax\maketitle}
\thispagestyle{empty}
\vskip 1.5cm
\begin{abstract}
We analyze observational signatures arising from an epoch in cosmology corresponding to the simple harmonic universe, which consists of positive curvature, a negative cosmological constant, and one or more exotic matter sources with $-1 < w = p/\rho \leq -1/3$, which may then evolve or tunnel into an inflating and hot Big Bang phase.  Particular relics specific to this class of models include spatial curvature with $\Omega_k < 0$, modifications to the dark energy sector due to the presence of residual exotic matter, and the possibility that the cosmological constant may ultimately be negative.  We review the constraints on the relevant parameters from Planck and other cosmological data sets, focusing on the roles of exotic matter and curvature, and we find that while sources with $w \approx -1$ tend to tighten the constraints on $\Omega_k$, matter with $w = -1/3$ can decouple curvature from the expansion history.  For the case where the dark energy sector includes a negative cosmological constant, we use the experimental data to find an associated lower bound on the lifetime of the expansion of the Universe. 
\end{abstract}

\section{Introduction}

The past several decades have seen a wealth of precision data for the study of early universe cosmology, and current and planned surveys of the cosmic microwave background (CMB) radiation and large scale structure (LSS) have the potential to test or rule out large classes of models and wide ranges of parameter space.  It is of particular interest to constrain or detect observational parameters that can probe the primordial era of cosmic inflation or even earlier, such as the tensor to scalar ratio $r$, or primordial non-Gaussianities, or spatial curvature $\Omega_k$.  This last parameter, if observable at a level greater than $10^{-5}$, may provide a rare window on the physics of the preinflationary epoch, such as whether our present observer patch may have originated from a tunneling event \cite{Coleman:1980aw, Guth:2012ww, Vardanyan:2009ft, Bull:2013fga}.  Furthermore, the sign of spatial curvature will have important consequences for model building \cite{Guth:2012ww, Freivogel:2005vv, Kleban:2012ph}.

In this paper we will compare the predictions of the class of models known as the simple harmonic universe (SHU), recently studied in \cite{Graham:2011nb, Graham:2014pca}, against experimental data from the CMB and from large scale structure.  The simple harmonic universe is a class of oscillating cosmological models which respect the null energy condition and which consist of a mixture of positive spatial curvature, a negative cosmological constant, and a matter source with $-1 < w = p/\rho < -1/3$ and positive sound speed squared.  In \cite{Graham:2011nb, Graham:2014pca} it was shown that these models can be stable at the level of linearized perturbations.  While the original motivation for studying the SHU class of models was to investigate singularity theorems and the stability of bouncing cosmologies, given the possibility that it may not be eternally stable against nonperturbative processes (see also \cite{Mithani:2011en, Mithani:2014toa, Horn:2017kmv}), it is also of interest to investigate whether this class of models can evolve into our present Universe, and the consequences for experiments if so.  In particular, the specific ingredients of the SHU may give rise to distinctive relics, which may motivate novel search templates.

One such possible relic motivated by this class of models is positive spatial curvature with $\Omega_k < 0$\footnote{By convention, $\Omega_k < 0$ corresponds to positive (i.e., spherical or closed) spatial curvature, and $\Omega_k > 0$ corresponding to negative (hyperbolic or open) spatial curvature.}: in \cite{Horn:2017kmv} it was shown that the SHU may evolve into an era of inflation (followed by reheating and the hot Big Bang era) via nonperturbative quantum tunneling in the matter sector, while preserving the original spherical geometry.  While many theoretical models of the landscape predict $\Omega_k \geq 0$ to be more generic (\cite{Freivogel:2005vv, Kleban:2012ph}), in light of the existence of models that can produce $\Omega_k < 0$ as well (see also for instance \cite{Atkatz:1981tk, Vilenkin:1982de, Vilenkin:1983xq, Vilenkin:1984wp, Bousso:1998ed, Buniy:2006ed} for previous works on this subject), it seems worth revisiting the bounds on experimental searches for $\Omega_k < 0$. 

Another class of predictions specific to the SHU involves the exotic matter contributions, whether these leave a remnant that contributes to the present dark energy sector, and how the resulting dark energy sector may affect the limits on curvature.  A few scenarios specific to this class of models include a matter source with $w = p/\rho = -1/3$ that (almost) cancels the curvature energy (so that the geometric effects of curvature are decoupled from the expansion history), or a scenario where the dark energy sector consists of a \textit{negative} cosmological constant whose energy is canceled out by a matter sector with $w \gtrsim -1$.  In this case we may still be living in an oscillating phase, with the expansion of the universe destined to reverse in the future.   

This paper is organized as follows: in section 2 we briefly review the SHU class of models and the relevant components that may contribute observable relics if the SHU evolves or decays into the standard cosmological history of our present observer patch.  In section 3 we make use of data from the Planck satellite \cite{Ade:2015xua, Aghanim:2018eyx}, together with the BOSS DR12 baryon acoustic oscillation survey \cite{Alam:2016hwk} and the Pantheon supernova dataset \cite{Scolnic:2017caz}, to put limits on these parameters.  
We conclude and indicate further directions in section 4.
 
\section{Model and observables}

We start with the class of models known as the simple harmonic universe: the background solutions were developed in \cite{Graham:2011nb, Harrison1967, Kardashev1990, Dabrowski1996} and consist of a mixture of positive spatial curvature, negative cosmological constant, and a null energy condition (NEC) preserving matter source with $-1 < w = p/\rho < -1/3$.  Note that postive spatial curvature contributes as negative energy when written on the right hand side of the Friedmann equation; this may be partially or almost completely cancelled by an additional source with $w = -1/3$, which serves to help with the stability of linearized perturbations \cite{Graham:2014pca}.  The evolution of the scale factor $a(t)$ is then determined by the Friedmann equation
\begin{equation}
H^2 = \left(\frac{\dot{a}}{a}\right)^2 =  -\frac{K_{eff}}{a^2} + \frac{8\pi G_N}{3}\left(\rho_0\frac{a_0^{3(1+w)}}{a^{3(1+w)}} - \Lambda\right)\,,
\end{equation}
where $K_{eff} = 1 - (8\pi G_N \rho_{string} a_0^{2})$ consists of curvature and a matter source which scales energetically like a network of cosmic strings ($w = p/\rho = -1/3$), $\Lambda$ refers to the negative cosmological constant (c.c.), and $\rho_0$ is the energy density of the exotic matter source with intermediate equation of state $-1 < w = p/\rho < -1/3$.  Note that negative energy is required to reverse the sign of $\dot{a}$; here the requisite negative energy sources are the curvature and the c.c., both of which respect the null energy condition.  The negative curvature energy causes the universe to bounce at $a = a_{min}$, and the negative c.c. causes the universe to recollapse at $a = a_{max}$.  Depending on the relative contributions of the various energy sources, there may be a large hierarchy between the maximum and minimum values of the scale factor, or the universe may be in a nearly static limit with $a_{max}/a_{min} \approx 1$.  We do not consider in detail what the initial conditions that lead to the the SHU may be, but rather we note that once in this phase, this class of solutions can be stable on long timescales\cite{Graham:2011nb, Graham:2014pca}.

Although the original motivation of \cite{Graham:2011nb, Graham:2014pca} for exploring this class of models was to consider the stability of oscillating universes, it is also of interest to explore embedding the SHU into a more realistic cosmology, with the (perhaps long-lived) oscillatory phase evolving into an epoch of inflation, followed by reheating and radiation and matter domination.  The SHU may evolve smoothly into an inflating and hot Big Bang phase, or the transition may be mediated by a nonperturbative tunneling event as in \cite{Horn:2017kmv}.  We will not focus on the details of the model building at present, but rather indicate the general features of what such models would entail, and what observable relics may persist into the present epoch.  

\begin{itemize}

\item 1) If the universe evolves smoothly between the oscillating phase and an inflating one, then epochs of inflation, reheating, radiation and matter domination could be included as a transient part of a much longer overall bouncing cycle.  Inflation could even be driven by the exotic fluid itself if $w \approx -1$, with most or all of the fluid energy decaying into radiation at reheating.  The bounce may have happened only a few times, with some of the exotic matter sector being lost each time, or perhaps many times, with a finite but small probability of the exotic matter decaying into radiation and reheating the Universe during each cycle.  Either way, the exotic matter source may decay completely, or a remnant could persist until the next cycle.  In the latter case the dark energy today may consist of a fluid with $w > -1$, together with a negative cosmological constant that becomes apparent at very late times.

A slightly more complicated variant of this model might be for inflation to instead be sourced using a vacuum energy sector that depends on the value of a scalar field whose energy moves between positive values during inflation and a negative value at $-\Lambda$ as the background oscillates between its minimum and maximum scale factor.  In this case, however, it seems difficult for the scalar field to move back and forth over multiple such bounces without introducing considerable fine-tuning. 

\item 2) Another possibility is that the universe could exit the SHU oscillating phase by quantum tunneling into an inflating phase with positive curvature -- such a model was studied in \cite{Horn:2017kmv}, where the nonperturbative tunneling of a scalar field mediates the exchange of the energy of the exotic matter sector for the energy of the inflationary potential.  In this case the amount of energy in positive curvature may directly correlate with the energy scale of inflation, and as before, there may or may not be a remnant of exotic matter as well.

\end{itemize}

In either scenario, the residual relics from the SHU could include positive spatial curvature, with or without the negative energy canceled out by an accompanying contribution from cosmic strings, and/or a contribution to the dark energy sector from an exotic matter remnant with equation of state $-1 < w < -1/3$.  If the original negative c.c. is not uplifted during the transition to the inflationary epoch, then it may persist and cause the universe to recollapse at late times.  We will now discuss the observable effects of each of these relics in turn, starting with curvature and its interactions with the observables from exotic matter.  In the case where the dark energy sector includes a negative c.c., we can use the experimental bounds on cosmological parameters to predict a lower limit on the duration of the current expansion of the Universe.



\section{Observables and results}

In this section we will model the effects of curvature and exotic matter on precision cosmological observables, and investigate the experimental constraints on the relevant parameters.  We will make use of the Cosmic Linear Anisotropy Solving System (CLASS) \cite{Lesgourgues:2011re, Blas:2011rf} to model the evolution of the CMB and matter anisotropies, with slight modifications to the code to incorporate additional fluid sources as well as modifications to the primordial power spectrum.  To compare with experimental data sets and infer bounds on cosmological parameters, we make use of the MontePython cosmological parameter inference code \cite{Audren:2012wb, Brinckmann:2018cvx}.  Many parameters of interest can be studied using the CMB data alone, combining the temperature and lensing data to break the degeneracy between curvature and dark energy parameters \cite{Ade:2015xua, Aghanim:2018eyx}).  We make use the Planck likelihoods for high multipole, low multipole, and lensing data.  We will use the Planck 2015 likelihoods since the 2018 likelihoods only recently became available; however, we do not expect the bounds on the observables we are interested in to be appreciably affected by the change in the optical depth likelihood between 2015 and 2018, and it is simple to check this by analyzing the correlations in the data and incorporating the new priors.  To gain further precision and to help further isolate the effects of curvature, we combine the CMB data with additional data from the Baryon Oscillation Spectroscopic Survey of SDSS-III (BOSS) \cite{Alam:2016hwk} and the Pan-STARRS Supernova cosmology data (Pantheon) \cite{Scolnic:2017caz}.  We also use CLASS to study the effects of varying cosmological parameters on the matter power spectrum.


For a review of how varying various cosmological parameters affects the evolution and appearance of the CMB peaks, see e.g. \cite{Hu:2008hd}.  For the observable relics we will consider, namely modifications to the curvature and dark energy sectors, the dominant effects of positive curvature $\Omega_k < 0$ include pushing the acoustic peaks to lower multipoles due to large-scale lensing, as well as enhancements of the temperature power spectrum at low multipoles due to the late-time integrated Sachs-Wolfe (ISW) effect.


\subsection{Positive curvature}

\begin{figure}
\begin{center}
\includegraphics[scale=0.4]{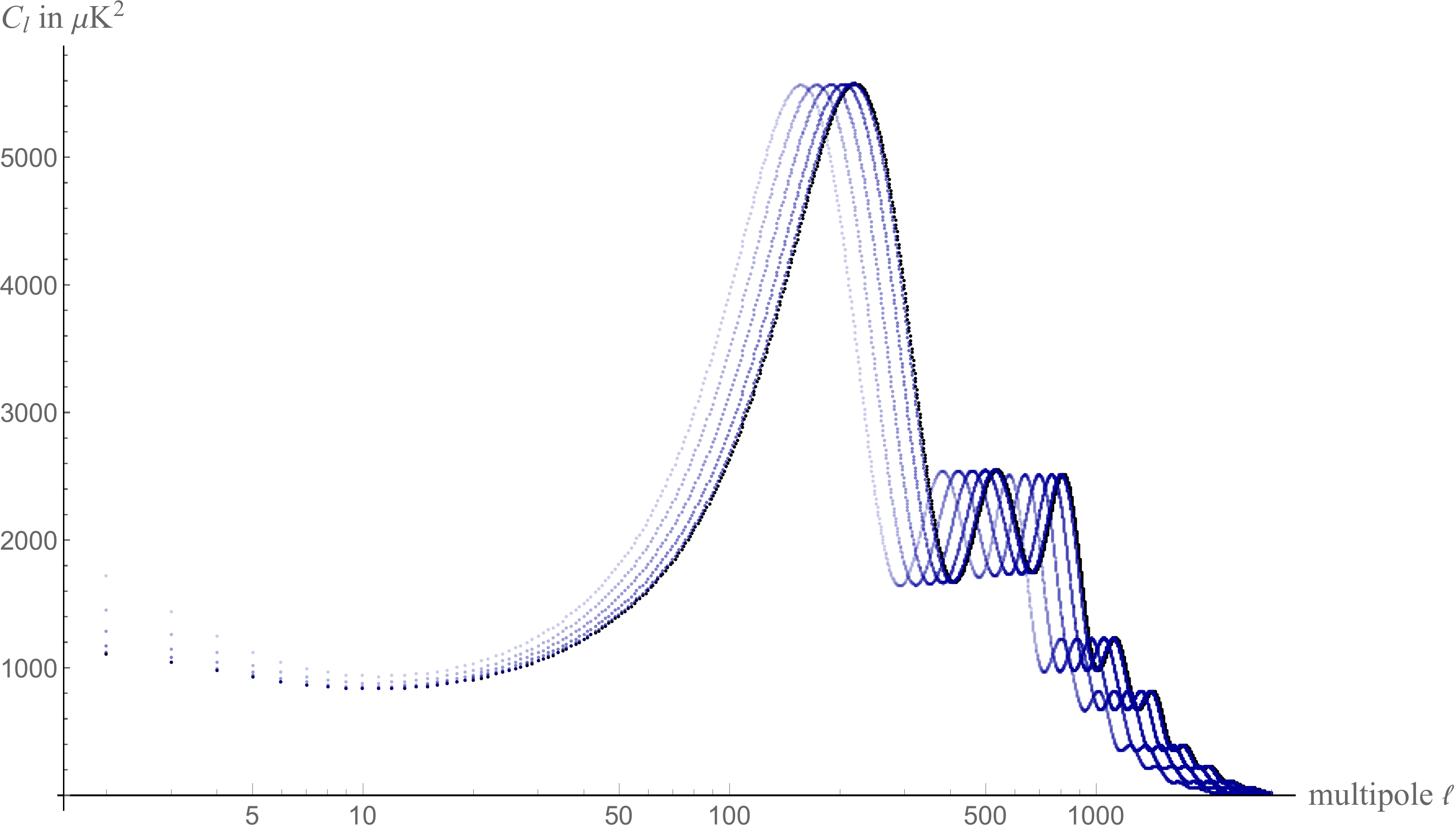}
\end{center}
\caption{Positive curvature and the CMB TT peaks.  Starting with $\Lambda CDM$ with background parameters of $\Omega_b = 0.05$, $\Omega_{cdm} = 0.26$, $\Omega_k$ is varied from $0$ (black) to $-0.20$, with the values $-0.01$, $-0.05$, $-0.10$, $-0.15$, and $-0.20$ shown in progressively lighter shades of blue.  Increasingly negative $\Omega_k$ tends to move the acoustic peaks to the left due to large-scale lensing, and the signal at low multipoles is enhanced due to the late-time ISW effect.
}\label{Omega_k}
\end{figure}

We will first review the constraints on spatial curvature, with special attention given to the case where the curvature is positive (spherical), which corresponds to $\Omega_k < 0$. See \cite{Leonard:2016evk} for discussion of the systematics and degeneracies involved in constraining $\Omega_k$, and see e.g. \cite{Okouma:2012dy, Takada:2015mma, Rana:2016gha, Witzemann:2017lhi} for model-independent searches for this parameter.  In the presence of positive spatial curvature, given a fixed physical sound horizon for the baryon acoustic oscillations, the lensing of the CMB acoustic peaks by the late-time geometry tends to move the peaks towards smaller multipoles, as shown in Fig.~\ref{Omega_k}.  There are also modifications at low multipoles due to the enhanced late-time ISW effect, since the presence of negative energy from positive curvature causes the Universe to deviate from matter domination earlier in its cosmic history.
Current observations constrain $|\Omega_k|$ at the one percent level using the CMB data alone, or at the level of $5 \times 10^{-3}$ when additional data sets (such as come from baryon acoustic oscillation or measurements of the Hubble parameter as a function of redshift) are included.  The precise $1\sigma$ limits reported by the Planck collaboration for $\Lambda CDM + \Omega_k$ are\footnote{The constraints here are those reported in \cite{Aghanim:2018eyx}, using chains generated by the Planck Collaboration.  For all other analyses we have rerun the chains ourselves using the Planck and other data sets.} \cite{Aghanim:2018eyx}:
\begin{equation}
\begin{split}
\Omega_k = -0.0106 \pm 0.0065 \qquad \textrm{(Planck only)}\\
\Omega_k = 0.0007 \pm 0.0019 \qquad \textrm{(Planck + BAO)}
\end{split}
\end{equation}
Note that $\Omega_k  < 0$ is slightly favored by the Planck data alone: this is mostly caused by the data at low multipoles, which are strongly affected by cosmic variance and by the lensing reconstruction.  This may ultimately be resolved in favor of a detection of of positive curvature (see e.g. \cite{DiValentino:2019qzk}); however, even without additional data sets the error bars are still consistent with a flat universe at the $2\sigma$ level, and including BAO data the error bars are consistent with a flat universe \cite{Ade:2015xua, Aghanim:2018eyx}.

\begin{figure}
\begin{center}
\includegraphics[scale=0.5]{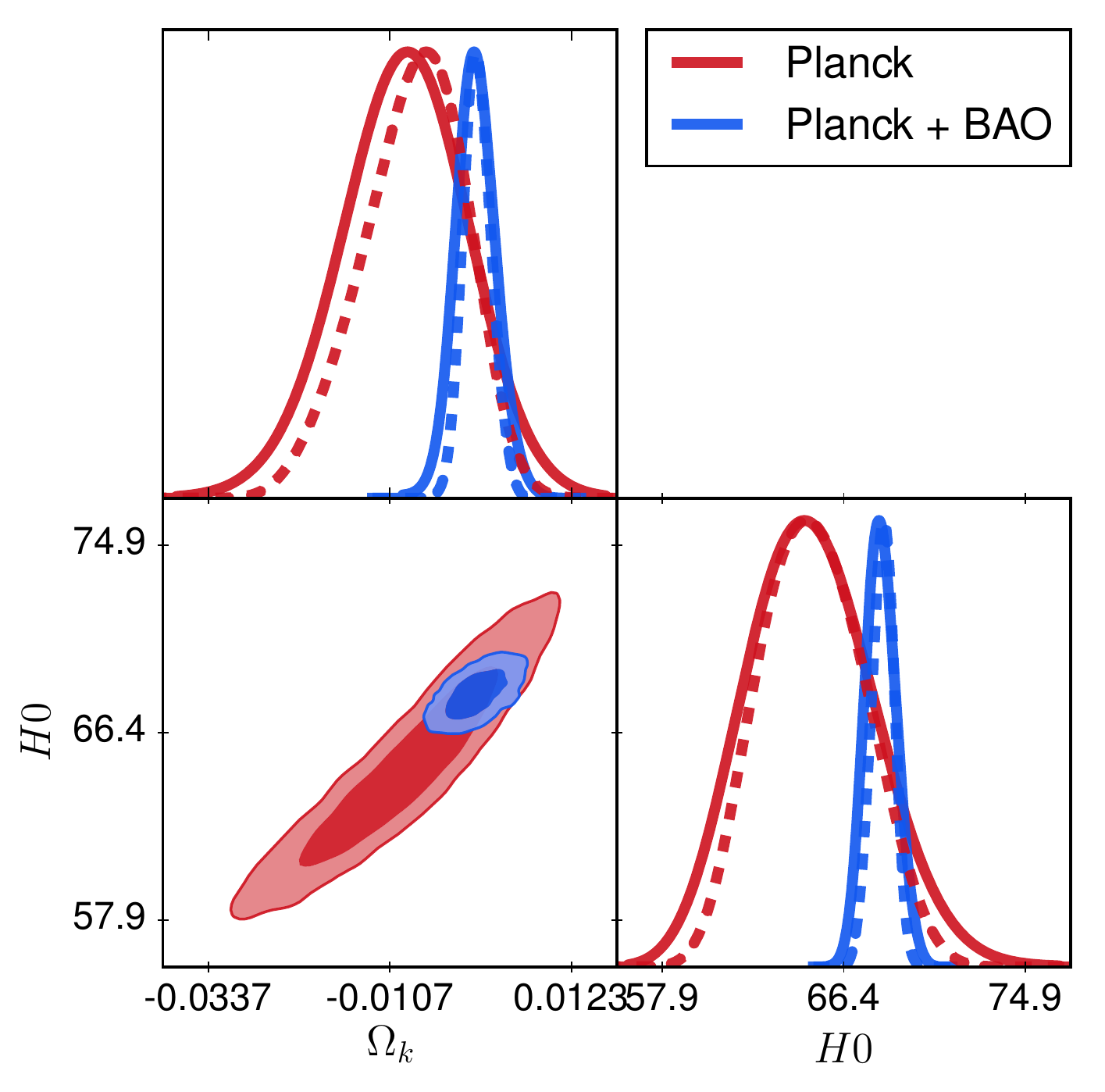}
\end{center}
\caption{2d likelihoods for $\Omega_k$ versus $H_0$.  Red: Planck data alone; Blue: Planck + BAO (BOSS DR 12) data.}\label{OmegaKvsH0}
\end{figure}

\begin{figure}
\includegraphics[scale=0.5]{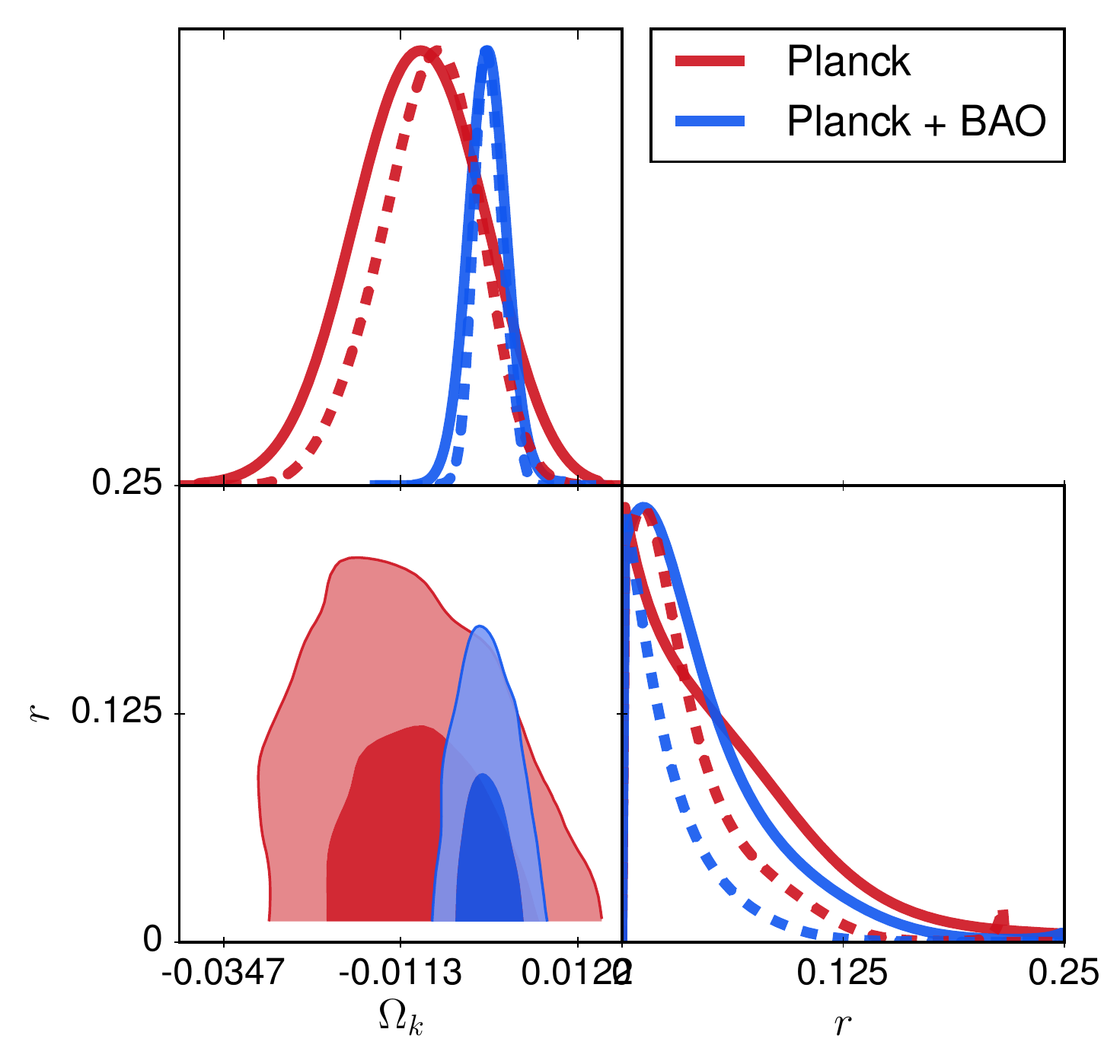}
\includegraphics[scale=0.5]{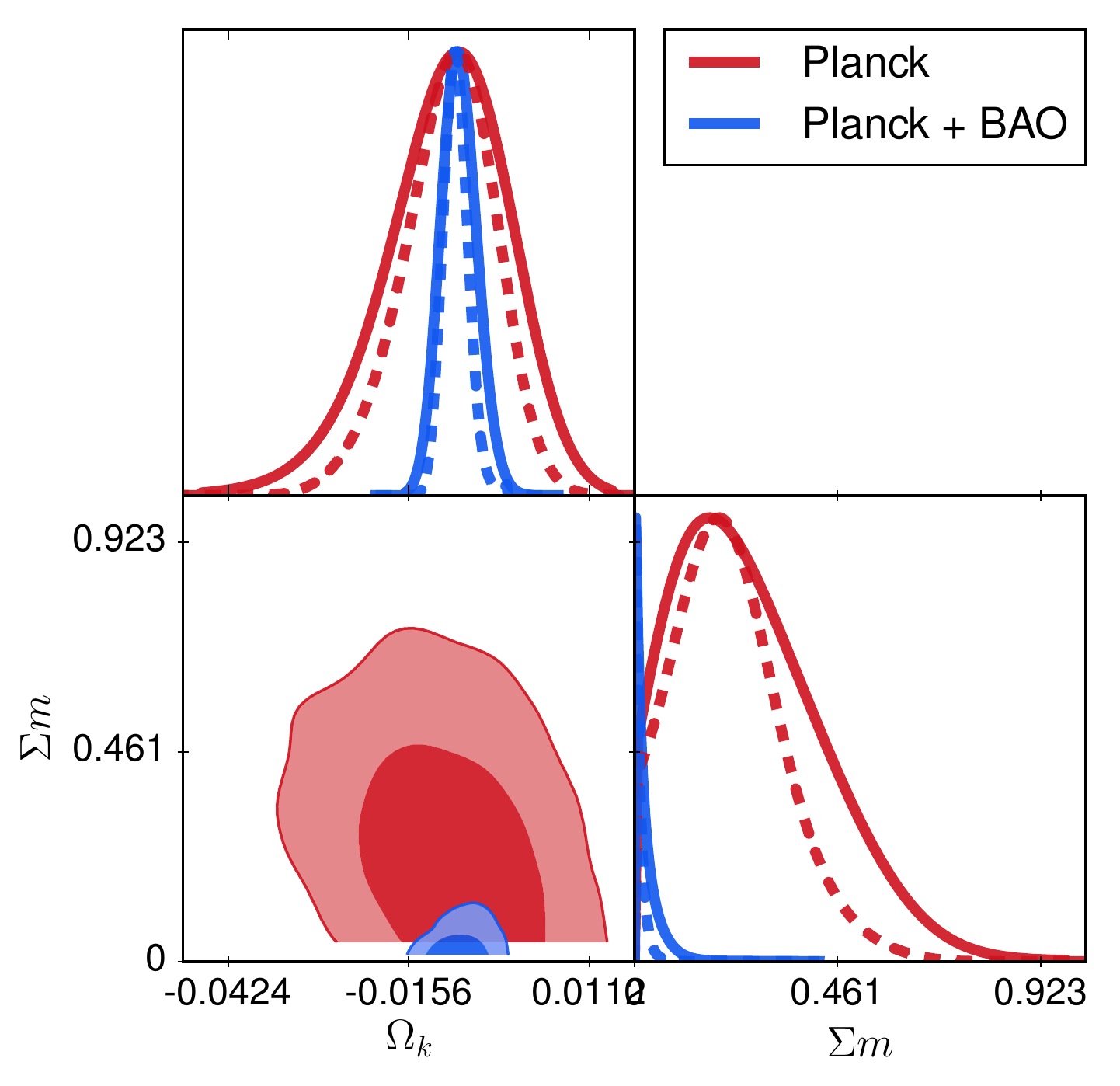}
\caption{Degeneracies between $1\sigma$ and $2\sigma$ likelihoods for $\Omega_k$, the tensor-to-scale ratio $r$, and the neutrino mass $\sum m_{\nu}$.  Red = Planck; Blue = Planck + BAO (BOSS DR 12) data.}\label{degeneracies}
\end{figure}

The bounds on curvature will be affected by the choice of dark energy model and on the value of $H_0$, since at the background level dark energy and curvature have similar effects on the CMB peaks (see figure \ref{OmegaKvsH0}).  
The degeneracy can be broken at the level of the Planck data by including lensing, and even more effectively by including BAO or other large scale structure data sets.  See e.g. \cite{Ooba:2017ukj, Ryan:2018aif, Park:2018tgj, Ryan:2019uor} for searches in the context of varying dark energy models.  The bounds will also exhibit some degeneracy with other observables which are affected by the lensing, such as the tensor-to-scalar ratio and the neutrino mass.  In Figure \ref{degeneracies} we plot the 2d likelihood contours when $\Omega_k$ is varied together with these parameters, checking against Planck and BAO data.  Note that the presence of a small positive curvature would relax the bounds on the tensor-to-scalar ratio very slightly (see figure \ref{degeneracies}), and also that the effects of positive curvature are slightly degenerate with the presence of massive neutrinos, as noted in \cite{Leonard:2016evk, Allison:2015qca}.  The degeneracy with massive neutrinos can be greatly reduced by including BAO data, but letting the neutrino mass vary does still shift the likelihood slightly towards $\Omega_k < 0$.

In a universe with positive curvature, the CMB data will also favor a slightly lower value of $H_0$, corresponding to an older Universe (see figure \ref{OmegaKvsH0}).
This tends to exacerbate the tension between CMB and late-time astrophysical measurements of this parameter \cite{Aghanim:2018eyx, Verde:2019ivm}; however, varying $\Omega_k$ by a percentage point in either direction is not enough to resolve this discrepancy anyway.  Additional effects such as massless radiation species, or dynamical dark energy, have been proposed to reconcile the different results for the Hubble parameter, and it is also possible that this will be resolved by a better understanding of the systematic errors (see e.g. \cite{Verde:2019ivm} for a discussion of the degeneracies and systematics involved).

While the tensor-to-scalar ratio is related to the Hubble scale during inflation, the implications of spatial curvature for the parameters of primordial era are more model-dependent.  
In general the relation between these parameters and the Hubble scale $H_{inf}$ is given by
\begin{equation}\label{relations}
r \sim 10^{10}\left(\frac{H_{inf}^{2}}{M_{P}^{2}}\right)\,, \qquad \Omega_k  \lesssim \left(\frac{a_{inf}H_{inf}}{a_0 H_0}\right)^2\,.
\end{equation} 
Here $a_{inf}, H_{inf}$ are the values of the scale factor and the Hubble parameter at the beginning of inflation, and $a_0 = 1, H_0 \sim 10^{-42}\textrm{GeV}$ are the values today.  The final $\lesssim$ will be an approximate equality if curvature (briefly) makes up an order one fraction of the energy density at the beginning of inflation, as arises e.g.\ in \cite{Horn:2017kmv}, where a tunneling event in the matter sector mediates a transition from the SHU to an inflating solution with positive curvature.  If both $r$ and $\Omega_k$ are detectable, it would therefore enable us both to probe both the value of the Hubble scale during inflation as well as to put a bound on the number of e-foldings between inflation and the present day.  There is considerable uncertainty in the theoretical prediction for $a_{inf}/a_0$, mostly coming from the end of inflation until reheating (see for instance \cite{Aslanyan:2015pma}), but at least 50-60 e-foldings are needed for inflation to solve the horizon and flatness problems, and at least a further 23 or so in order to reduce the temperature of the background radiation from $T_{BBN}$ to $2.7K$. 

\subsection{Curvature during inflation and the primordial power spectrum}

In addition to its effects on the CMB peaks at late times, if curvature is present during inflation it may have consequences for the generation of the primordial spectrum.  In particular the presence of spatial curvature will cut off the power spectrum for inflationary fluctuations with wavelengths larger than the curvature radius, which may account for the observed suppression of the TT quadrupole and other low multipoles in the Planck data  \cite{Efstathiou:2003hk}.  However, the exact amount of the suppression depends on the initial conditions for the state of the perturbations, which may deviate significantly from the Bunch-Davies state.  More precise calculations of the inflationary perturbation equations in the presence of curvature for several different choices of initial conditions were carried out in \cite{Ratra:2017ezv, Handley:2019anl}.  In \cite{Avis:2019eav} it was noted that the sensitivity to spatial curvature may be boosted considerably if the sound speed of the perturbations is small, in which case we also expect distinctive features in the bispectrum.

\begin{figure}
\includegraphics[scale=0.4]{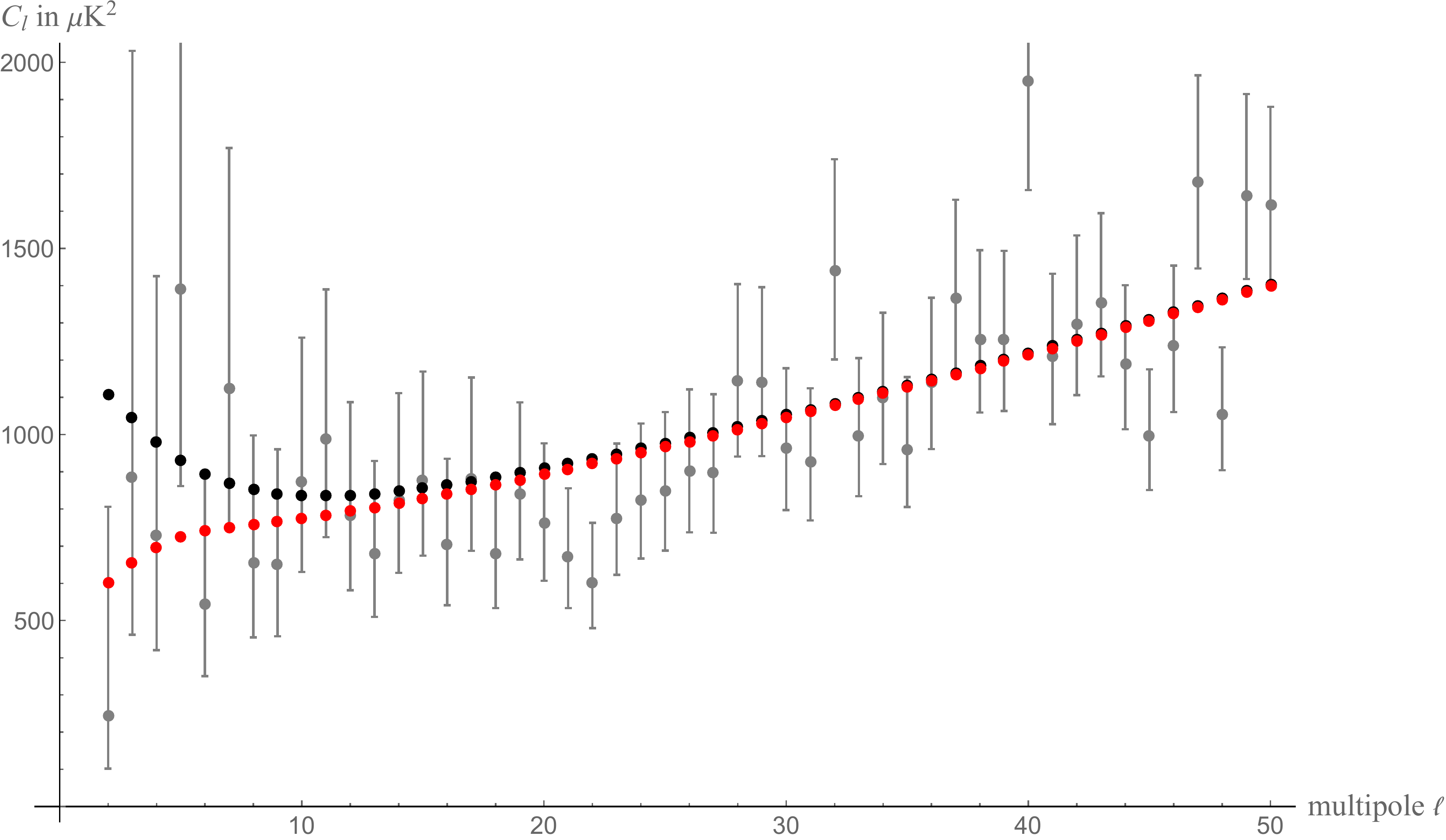}
\caption{The CMB TT power spectrum as a function of multipole, for the multipoles $\ell \lesssim 50$.  Shown are the best-fit $\Lambda CDM$ model with $k_0 = 0$ (black) and $k_0 = 2.0 \times 10^{-4} Mpc^{-1}$ (red).  Planck data and error bars are shown in grey.}\label{modLowL}
\end{figure}

We can test for these effects and the wavenumbers at which they emerge by taking the following simplified ansatz for the spectrum of inflationary perturbations:
\begin{equation}\label{powerspectrum}
P(k) \propto \left(\frac{k^2}{k^2 + k_0^2}\right)^{2}(k/k_{pivot})^{(n_s - 1)}\,.
\end{equation}
Here $(n_s - 1)$ is the scalar tilt, and the first factor cuts off the power spectrum at a scale $k_0$.  It would be interesting to derive a more detailed ansatz starting from a full model of the initial conditions; however, the ansatz here has the qualitative features we expect and the scale of the cutoff should not depend too sensitively on the precise functional form.  We can plot the effects of $k_0$ on the low multipole data (see Fig.~\ref{modLowL}).
We find a best-fit value of $k_0 \sim 1.68 \times 10^{-4}$ inverse Mpc, which corresponds to a comoving length scale on the order of 6 Gpc for the curvature radius.  This value for the curvature radius would produce a significant amount of $\Omega_k$ today:
\begin{figure}
\begin{center}
\includegraphics[scale=0.5]{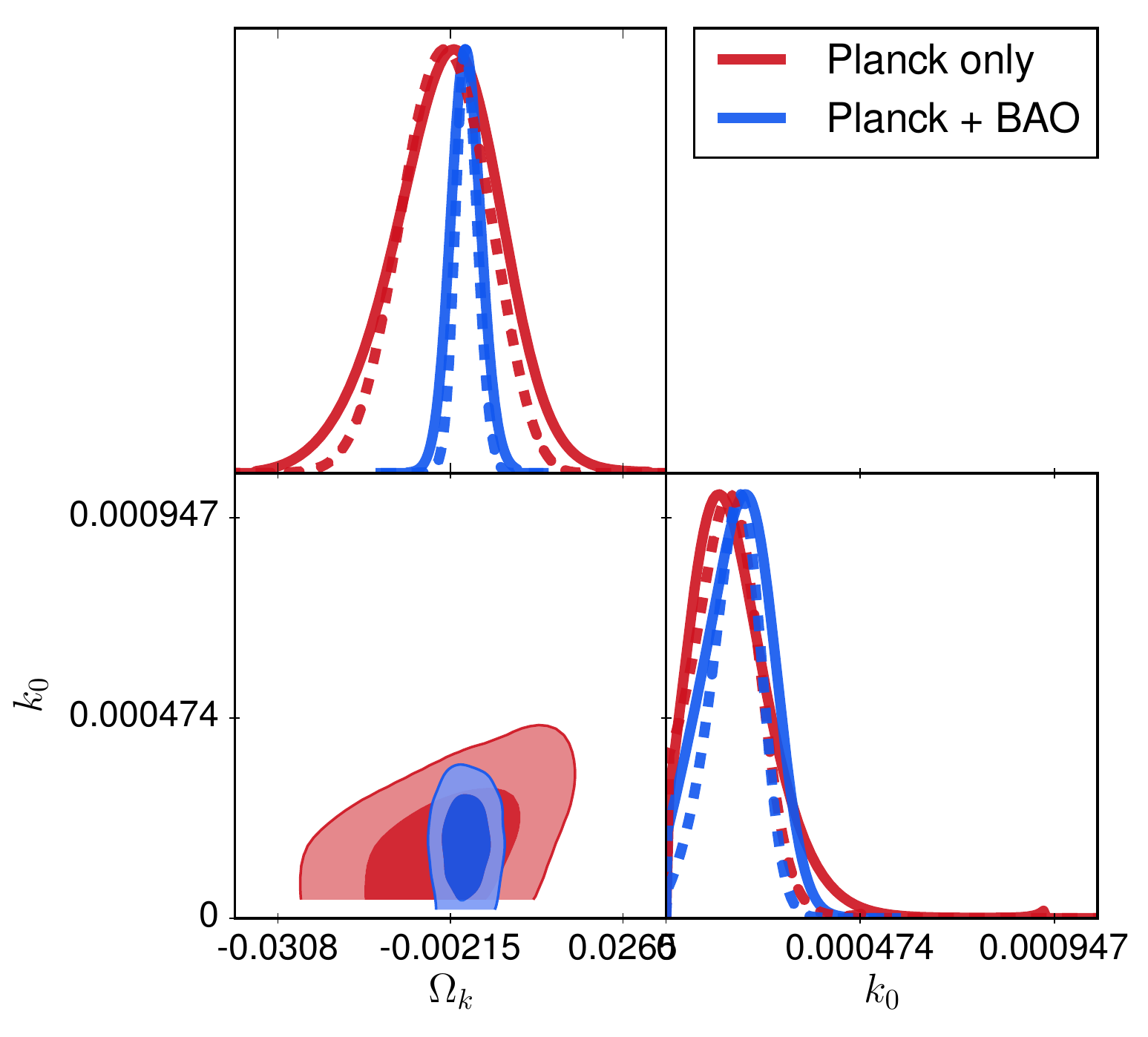}
\end{center}
\caption{$1\sigma$ and $2\sigma$ likelihood contours for the curvature parameter $\Omega_k$ and the scale $k_0$ of a low-wavenumber cutoff in the primordial spectrum.  Red = Planck data only, Blue = Planck + BAO (BOSS DR 12) data.}
\end{figure}\label{OmegaKvsk0}
\begin{equation}
\Omega_k \approx  -\left(\frac{k_0}{a_0 H_{0}}\right)^2 = -0.56\,.
\end{equation}
Since the experimental constraints of the previous section imply that $|\Omega_k| < 0.01$ or below, it may be that another explanation besides curvature is required for the suppression of the lowest TT multipoles; however, it is also possible that the observations could be reconciled if modes in the primoridal spectrum with wavelengths far below the curvature radius are sensitive to the curvature scale.  This can arise e.g. either through a small sound speed for the perturbations \cite{Avis:2019eav} or through orbifolding.

In the case of a small sound speed, the inflationary perturbations freeze out when they cross the sound horizon:
\begin{equation}
\frac{k}{a} \sim \frac{H}{c_s}
\end{equation}
and if $c_s \ll 1$, this may be much larger than the inverse radius of curvature.  A similar effect arises when a positively curved universe is orbifolded; for example, the shape of the spatial slice may be a Lens space orbifolded along the Hopf fiber rather than the full 3-sphere: $S^{3} \to S^{3}/\mathbb{Z}_m$.  In this case only spherical harmonics that are multiples of $m$ are allowed (an effect used in \cite{Graham:2014pca} to project out some of the possible perturbations), in which case the cutoff in the spectrum happens for modes with wavelengths a factor of $\mathcal{O}(\sqrt{m})$ or so smaller than the curvature radius. Reconciling the preferred value of $k_0$ with the bound $|\Omega_k| \lesssim 0.005$ therefore implies a bound $m \gtrsim 14$ on this class of models.  In principle the orbifolded models do not require strong self-interactions in the inflationary sector and need not give rise to a large bispectrum, making it potentially distinguishable from the models with small sound speed.

We emphasize that the ansatz for the power spectrum in \eqref{powerspectrum} is not derived from a specific model for the initial state, but rather serves to identify at what scale the spectrum must be cut off in order to explain the observed suppression of the low CMB multipoles.  It would be interesting to pursue this question further in context of a more detailed model of the initial state, and to investigate the effects on the bispectrum and higher-point correlation functions as well.


\subsection{Positive curvature and strings: $K_{eff} \ll 1$}

\begin{figure}
\includegraphics[scale=0.4]{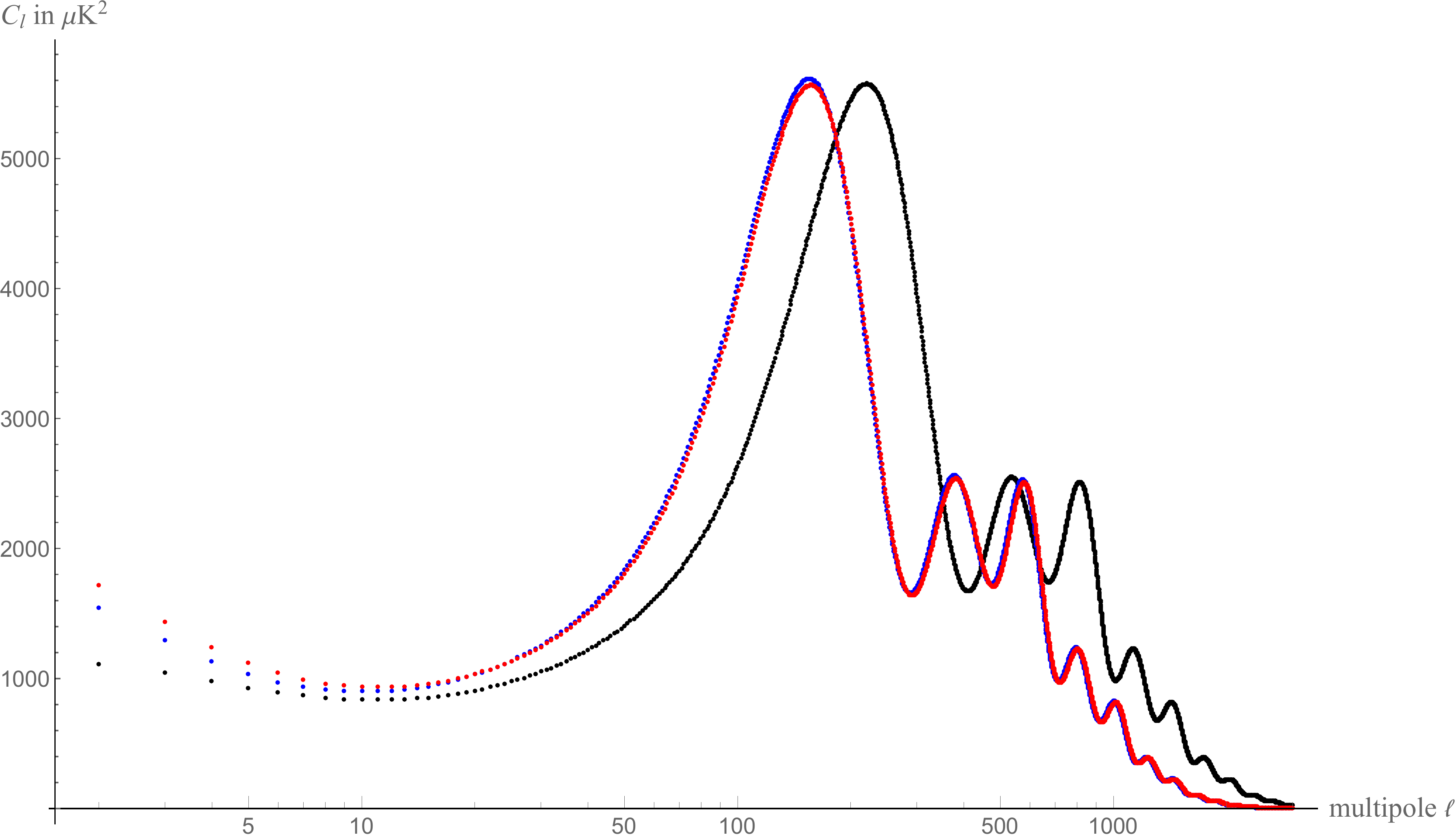}
\caption{Effects of $K_{eff}$ on the CMB TT peaks, in $\Lambda CDM$ with $\Omega_b = 0.05$, $\Omega_{cdm} = 0.26$.  Black: $\Omega_k = \Omega_{w = -1/3} = 0$, Red: $\Omega_k = -0.20, \Omega_{w = -1/3} = 0$, Blue: $\Omega_k = -0.20, \Omega_k + \Omega_{w=-1/3} = 0$.}\label{stringCheese}
\end{figure}

One possible corner of model space that arises in the context of the simple harmonic universe is the case where the negative energy from $\Omega_k < 0$ is almost totally canceled by a matter sector with $w = -1/3$ (which scales energetically the same way as a network of cosmic strings) and positive sound speed.  This corresponds to the case $K_{eff} \ll 1$ in the notation of \S 2.  In this case, the geometry of the universe will be affected by the curvature but not its late-time expansion history.  The CMB TT peaks are therefore translated slightly in multipole space relative to the corresponding model with $K_{eff} = 1$, and the low multipoles are slightly suppressed, though the CMB anisotropies are still larger than the model with $\Omega_k = 0$.  The effects on the anisotropies are plotted in figure \ref{stringCheese}.  

\begin{figure}
\includegraphics[scale=0.4]{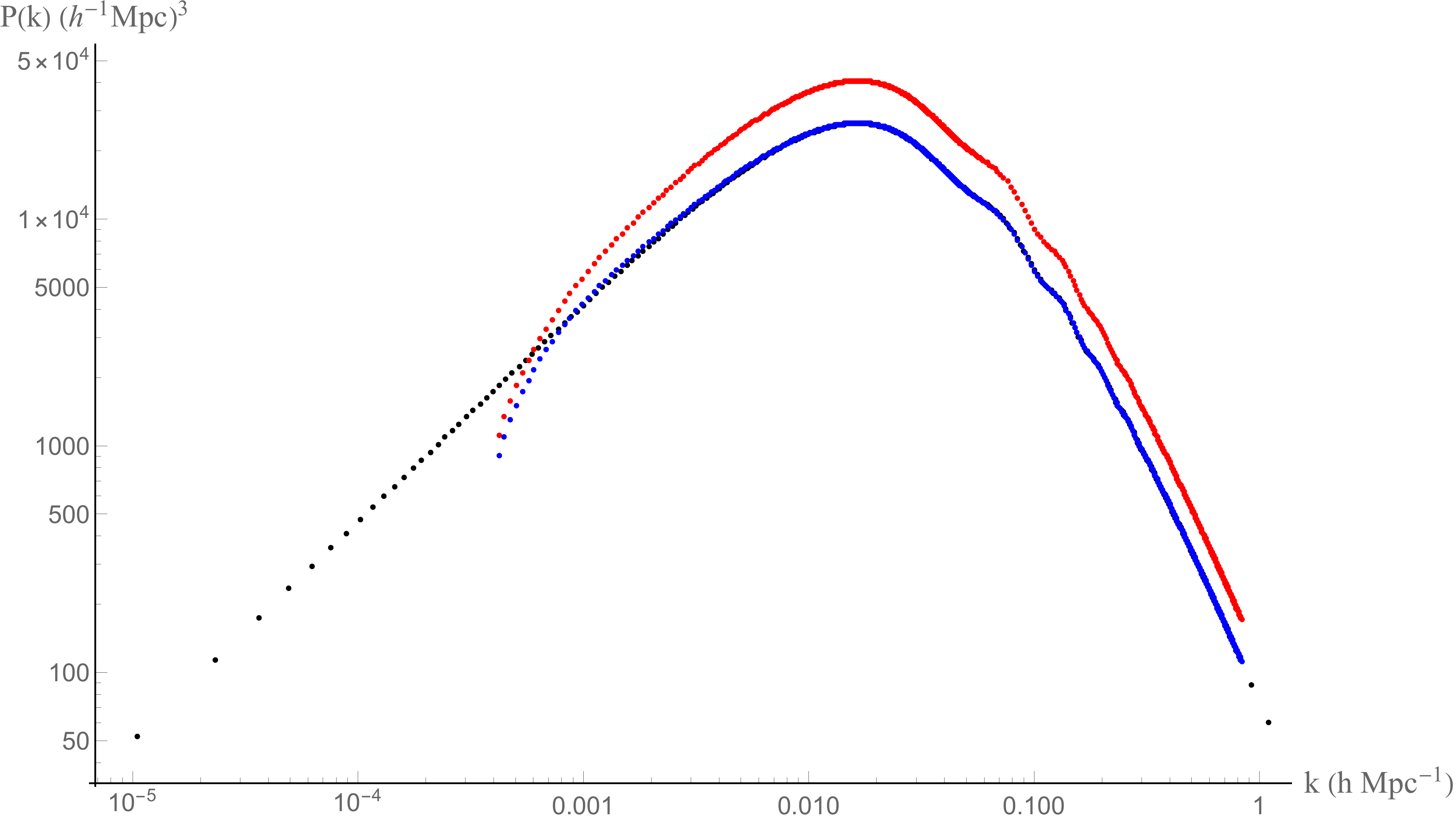}
\caption{Effects of $K_{eff}$ on the matter power spectrum peaks, in $\Lambda CDM$ with $\Omega_b = 0.05$, $\Omega_{cdm} = 0.26$.  Black: $\Omega_k = \Omega_{w = -1/3} = 0$, Red: $\Omega_k = -0.20, \Omega_{w = -1/3} = 0$, Blue: $\Omega_k = -0.20, \Omega_k + \Omega_{w=-1/3} = 0$, 
}\label{matter_power_curvature}
\end{figure}

The effects of $K_{eff} \ll 1$ are more noticeable at the level of the matter power spectrum: this is suppressed at very low wavenumbers due to the finite size of the closed universe.  Note that when positive curvature is not decoupled from the expansion history ($K_{eff} = 1$), the power spectrum is enhanced since the age of the Universe increases (see figure \ref{matter_power_curvature}).

\begin{figure}
\begin{center}
\includegraphics[scale=0.5]{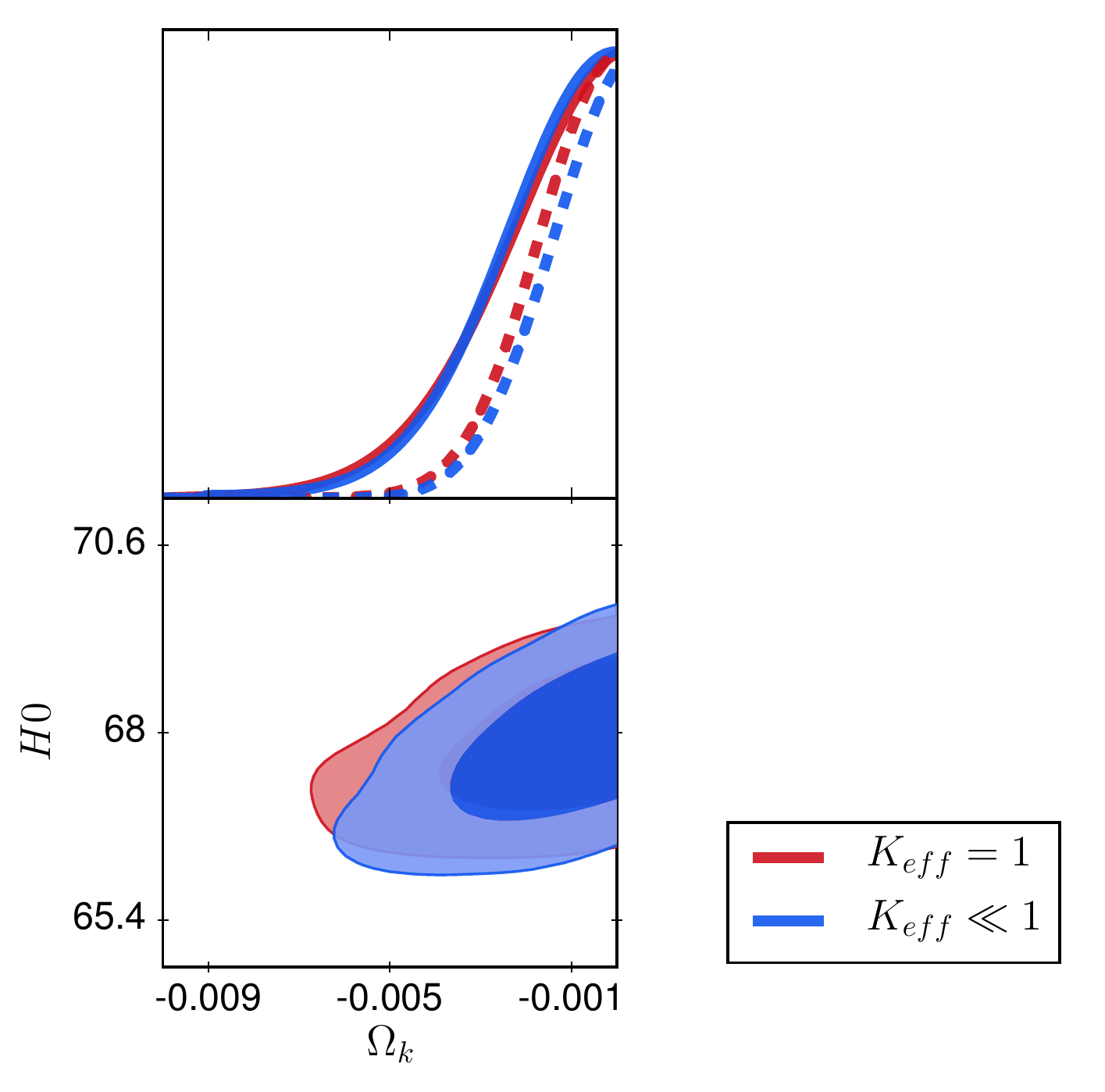}
\end{center}
\caption{Effects of varying $K_{eff}$ on the Planck + BAO (BOSS DR 12) likelihoods for $\Omega_k$ versus $H_0$.  Red = $K_{eff} = 1$, Blue = $K_{eff} \ll 1$.}\label{stringcheeseparams}
\end{figure}

To find bounds on the cosmic parameters, we consider the Planck data together with BAO data.
Marginalizing over the other parameters in $\Lambda CDM + r + n_s + \Omega_k$, the window of allowed values of $\Omega_k$ shrinks slightly, and the window of allowed values of $H_0$ is slightly expanded.  The relative likelihoods between $\Omega_k$ and $H_0$
is shown in figure \ref{stringcheeseparams}: as expected, the correlation between these parameters decreases slightly relative to the case with $K_{eff} = 1$.  
This helps prevent positive curvature from exacerbating the tension between CMB and astrophysical measurements of $H_0$, but not nearly enough to resolve the overall tensions.  

\subsection{Modified equation of state for dark energy}

In addition to looking for curvature alone, we can also search for the effects of modifications to the dark energy sector, coming from the presence of exotic matter and c.c. left over from the SHU epoch.   Depending on the model under consideration the exotic matter may make up part or all of the dark energy sector, and the cosmological constant $\Lambda$ may be positive or negative depending on whether $\Lambda$ is also a relic from the SHU epoch or whether it is altered in the transition from the SHU to an expanding phase.  We parameterize the exotic matter by including an extra fluid $\Omega_{fld}$ in CLASS with $-1 < w_{0fld} < -1/3$, and positive sound speed squared $c_s^2 = 1$\footnote{More precisely, the perturbations are implemented in our Monte Python analyses by using the parameterized post-Friedmann approximation as in \cite{Fang:2008sn}; however, we emphasize that we are sticking to the region where $w_{0fld} > -1$ for the exotic fluid, so we do not expect this to make much of a difference.}.
\begin{figure}
\begin{centering}
\includegraphics[scale=0.3]{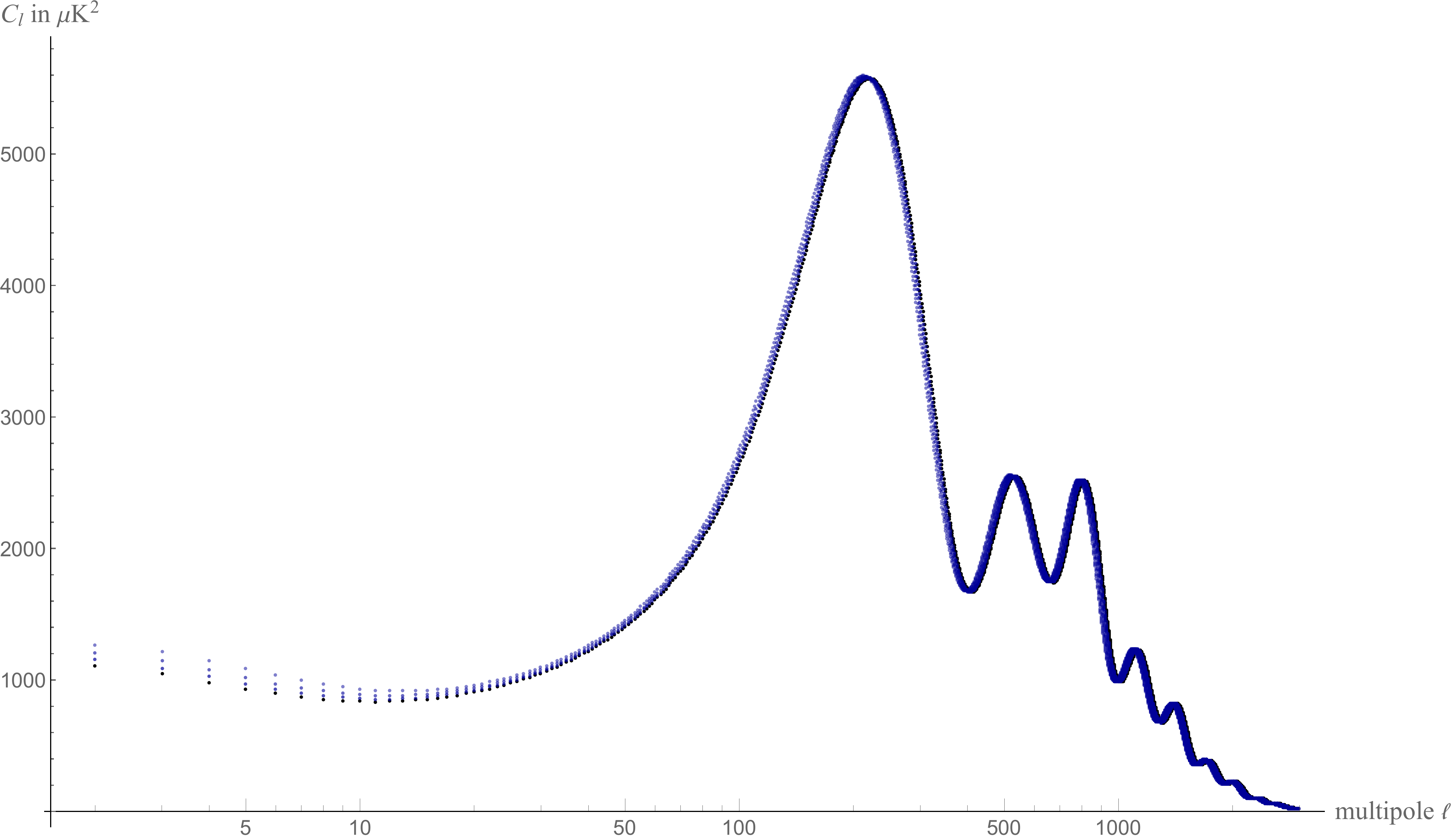}
\end{centering}
\caption{Exotic matter equation of state and the CMB peaks.  Starting with background parameters $\Omega_b = 0.05$, $\Omega_{cdm} = 0.26$, $\Omega_k = 0$, and $\Omega_{fld} = 0.69$ for the exotic matter sector, the value of $w_{0fld} = p/\rho$ for the exotic matter equation of state is varied from $-1$ to $-0.9, -0.8, -0.7$ (dark to light blue).  
}\label{modified_w}
\end{figure}

\begin{figure}
\begin{centering}
\includegraphics[scale=0.3]{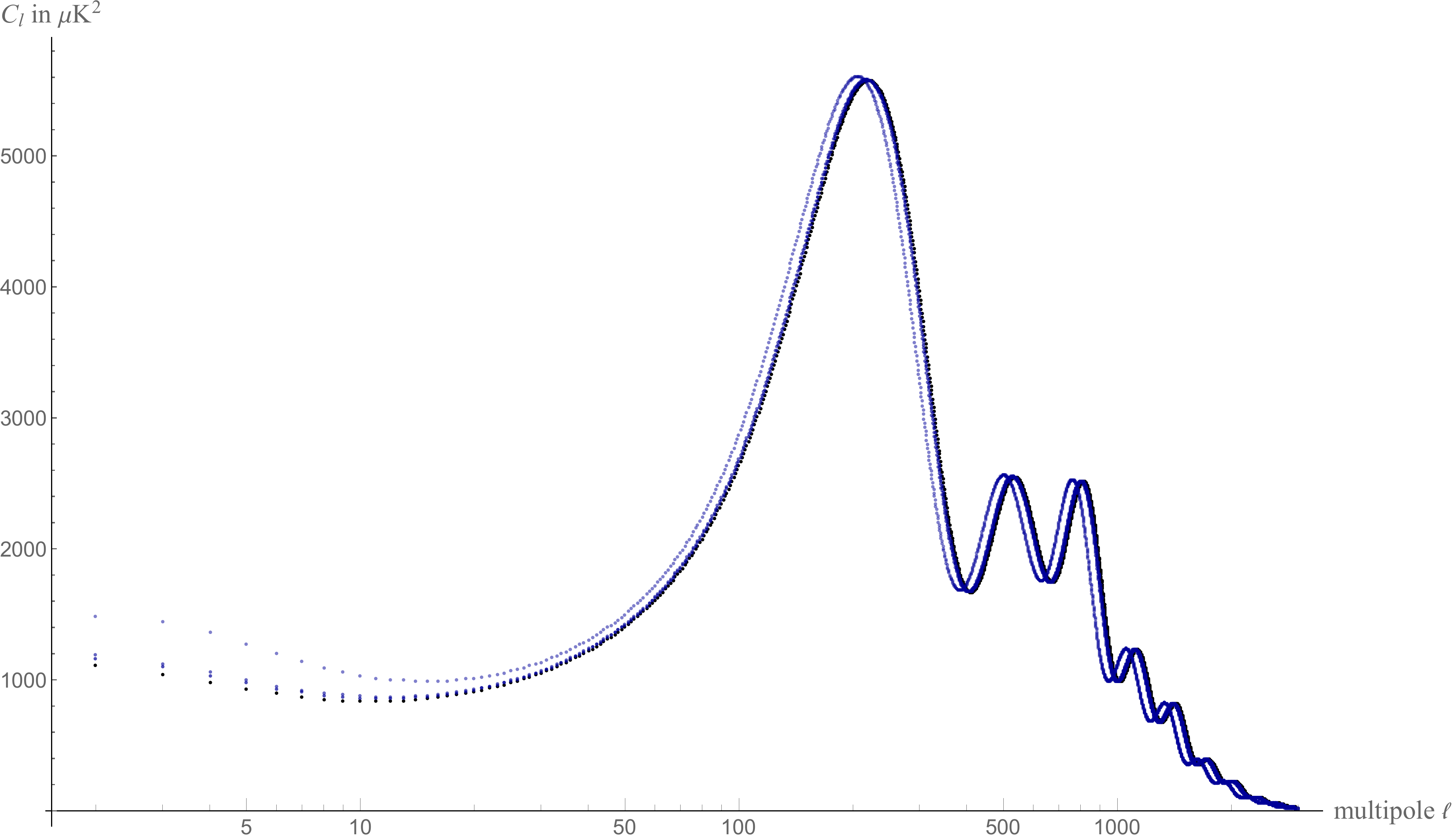}
\end{centering}
\caption{Exotic matter with $\Omega_{\Lambda} < 0$ and the CMB peaks.  Starting with background parameters $\Omega_b = 0.05$, $\Omega_{cdm} = 0.26$, $\Omega_{k} = 0$, and dark energy with $\Lambda < 0$ and dark fluid with $w_{0fld} = -0.9$.  Depicted: $\Omega_{\Lambda} = -0.1, -0.5, -5$ (dark to light blue).
}\label{modified_Omegafld}
\end{figure}
\begin{figure}
\begin{centering}
\includegraphics[scale=0.4]{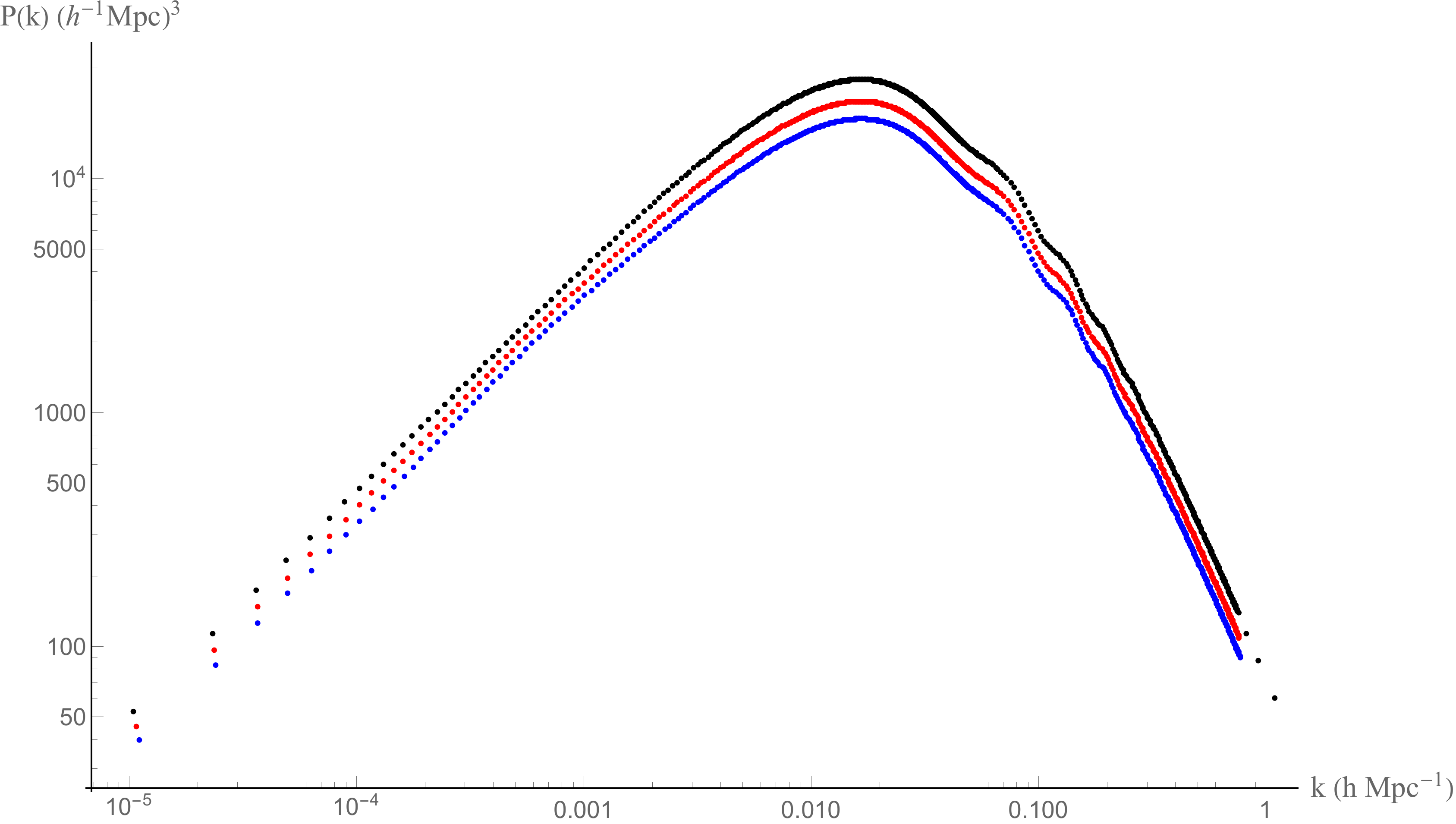}
\end{centering}
\caption{Matter power spectrum and modified dark energy: Black: $\Lambda CDM$ with $\Omega_b = 0.05$, $\Omega_{cdm} = 0.26$, $\Omega_k = 0$; Blue: $w = -0.7$; Red: $\Omega_{fld} = 5.68$, $\Omega_{\Lambda} = -5.0$ and $w_{0fld} = -0.9$.  
}\label{modified_DE_mPk}
\end{figure}
\begin{figure}
\begin{center}
\includegraphics[scale=0.5]{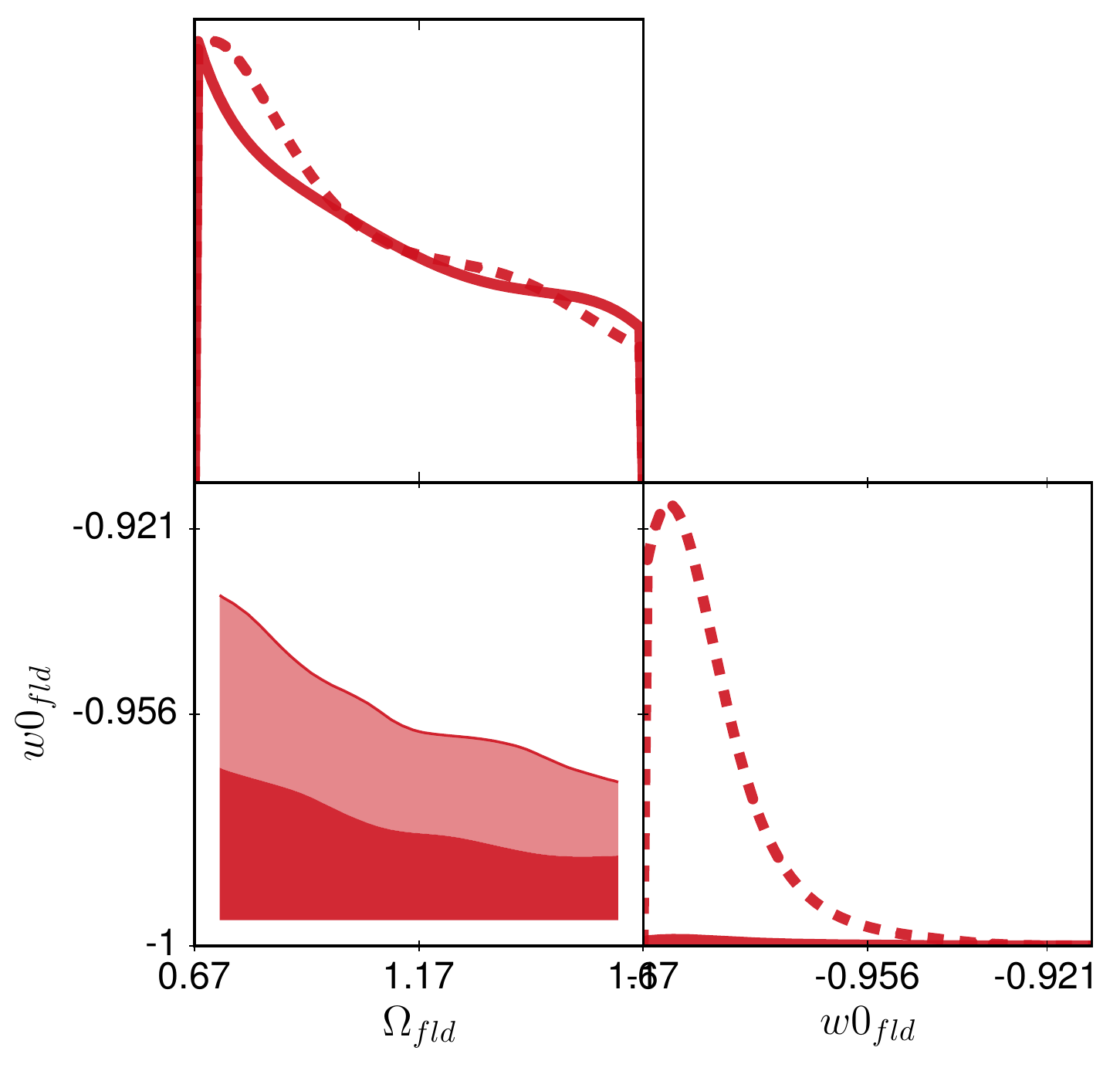}
\end{center}
\caption{$1\sigma$ and $2\sigma$ likelihood contours for the parameters $\Omega_{fld}$ and $w_{0fld}$ for exotic matter, including Planck + BAO (BOSS DR 12) + SNe (Pantheon) data sets.}\label{exotics}
\end{figure}

We can search for exotic matter in the dark energy sector by varying both the amount $\Omega_{fld}$ and the equation of state $w_{0fld}$ simultaneously.  The effects of varying each parameter separately on the CMB are modeled in figures \ref{modified_w} and \ref{modified_Omegafld} and are qualitatively similar: modifications away from $w = -1$ tend to shift the CMB peaks towards lower multipoles due to large scale lensing, and the power at low multipoles also increases due an enhanced late time ISW effect: both modifications mean that dark energy takes over from matter domination earlier in the history of the Universe.  The effects on the matter power spectrum are modeled in figure \ref{modified_DE_mPk}\: since dark energy takes over at an earlier epoch, the universe is younger and the matter power spectrum is suppressed relative to $\Lambda CDM$.  
When $w_{0fld} \approx -1$ we expect the constraints on $\Omega_{fld}$ to be weak, since the exotic matter is nearly indistinguishable from dark energy, but if $w_{0fld} > -1$ only a small amount of exotic matter will be allowed.  The Planck + BAO (BOSS DR 12) + SNe (Pantheon) constraints on $\Omega_{fld}$ and $w_{0fld}$ are plotted in figure \ref{exotics}, focusing on the region where $\Omega_{\Lambda} < 0$.  In order to keep the constraints as model-independent as possible; we have allowed the neutrino mass $\Sigma m_{\nu}$ to be arbitrary as well; however, this does not affect the bounds on $w_{0fld}$ versus $\Omega_{fld}$ appreciably.  

\begin{figure}
\includegraphics[scale=0.5]{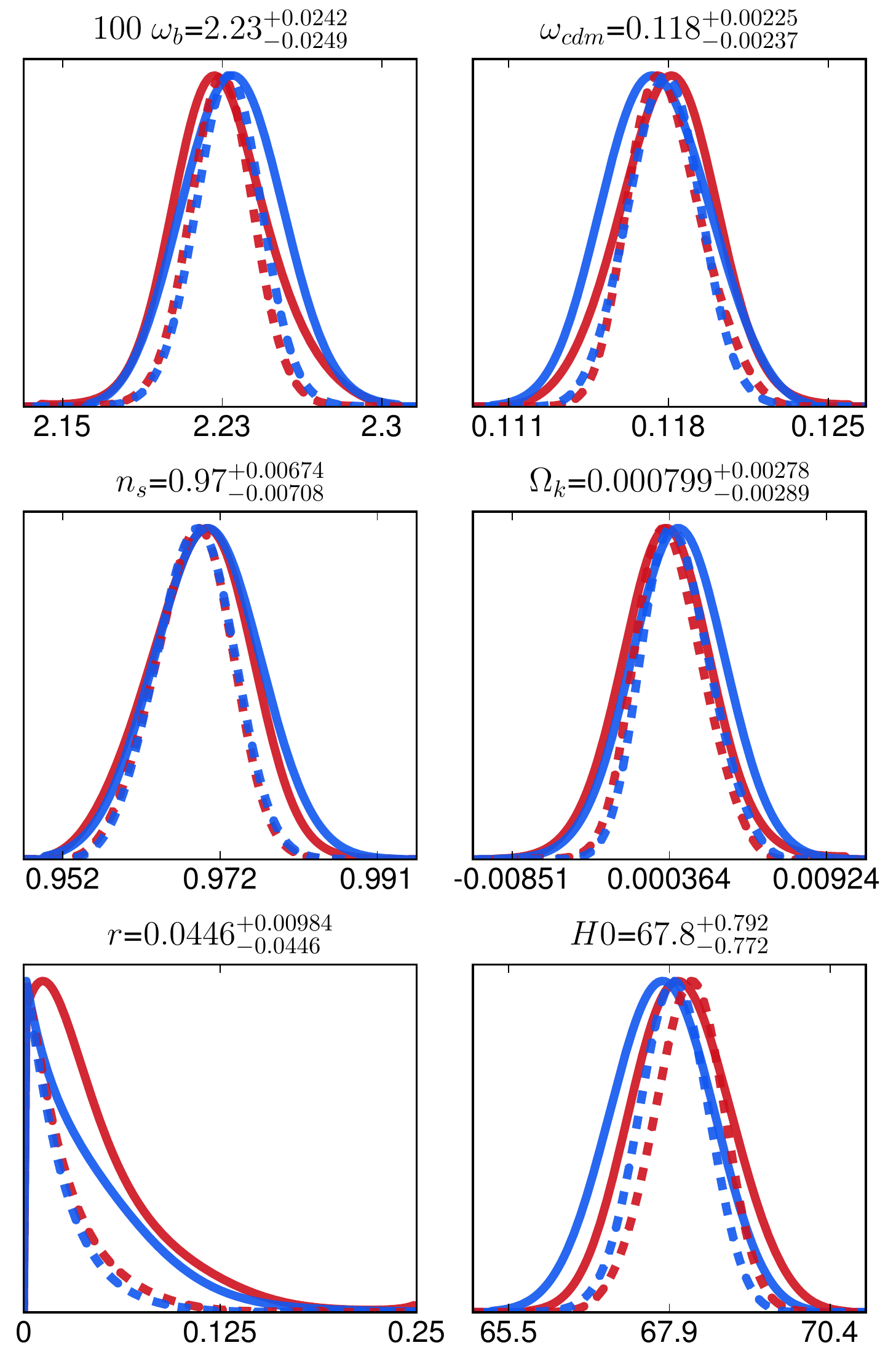}
\includegraphics[scale=0.5]{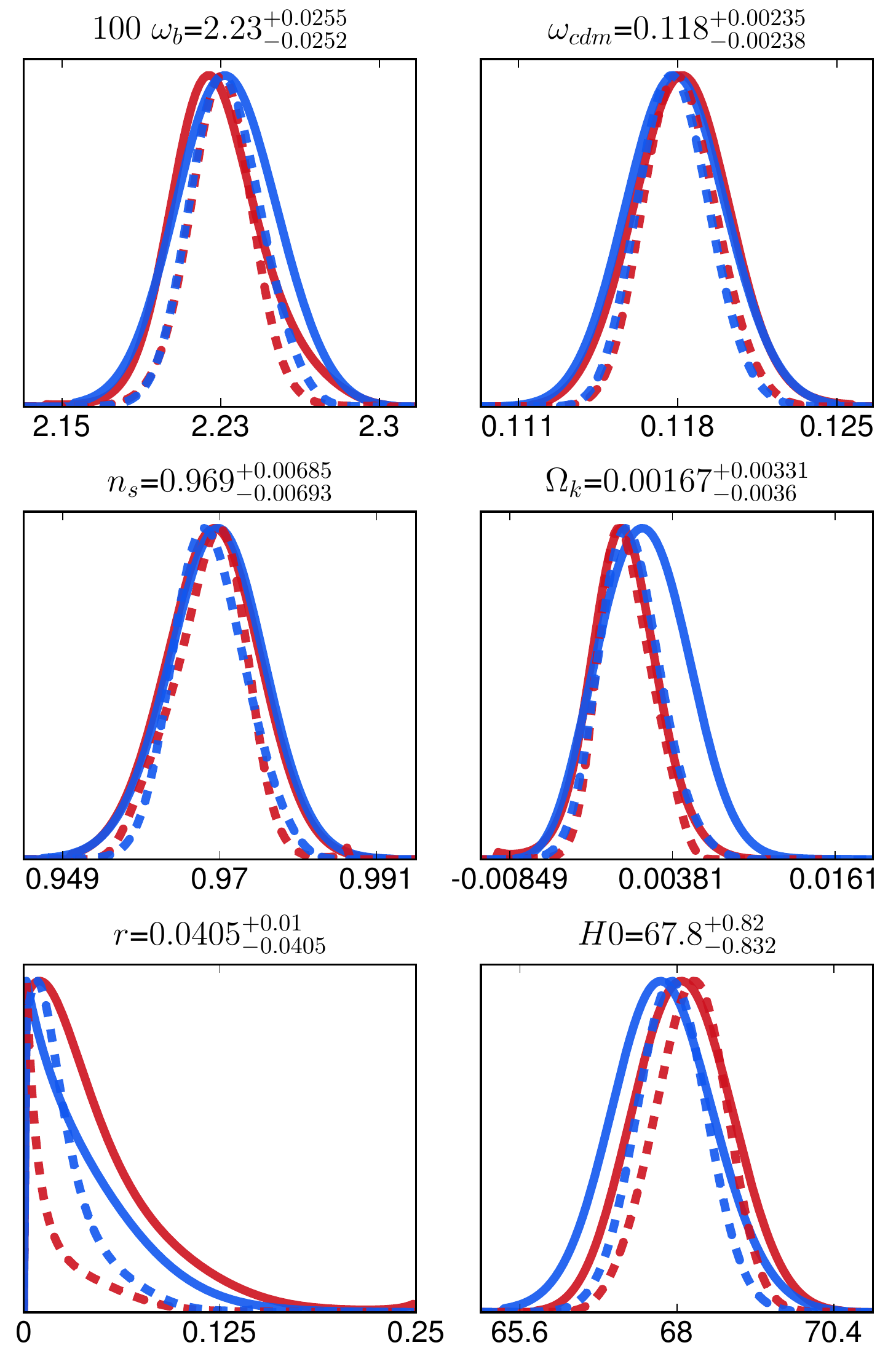}
\caption{Effects of modified dark energy with $\Omega_{fld}$ and $\Omega_{\Lambda} < 0$ on cosmological parameter likelihoods. Blue: Planck + BAO (BOSS DR 12) + SNe (Pantheon) data.  Left (blue): $\Sigma m_{\nu} = 0.06 eV$, Right (blue): $\Sigma m_{\nu}$ varying.  Red: $\Lambda CDM + n_s + r + \Omega_k$, with $\Sigma m_{\nu} = 0.06 eV$, using Planck + BAO data.}\label{OmegaKonly}
\end{figure}

Since both types of modifications to the dark energy sector push the acoustic peaks in the same direction(s) as does positive curvature, that is, the peaks move to the left and the power increases at low multipoles, including exotic matter in the dark energy sector tends to push the constraints on curvature towards $\Omega_k > 0$.  This is visible in figure \ref{OmegaKonly}, where including modified dark energy with $\Omega_{fld} > 0.67, w_{0 fld} > -1$ and $\Omega_{\Lambda} < 0$ shifts the error bars and best-fit value for $\Omega_k$ slightly towards the positive direction relative to $\Lambda$CDM.  The modified dark energy also favors a smaller Hubble constant, slightly worsening the tension with astrophysical estimates \cite{Aghanim:2018eyx, Verde:2019ivm}.  


Of the scenarios discussed in \S 2, note that in the case where inflation itself is sourced by exotic matter with $w_{0fld} \approx -1$, with most of the exotic matter decaying into the Standard Model at the epoch of reheating, then the slow-roll parameter will be
\begin{equation}\label{nstow}
\epsilon = -\frac{\dot{H}}{H^2} = \frac{3}{2}(1+w_{0fld})
\end{equation}
The Planck data are inconsistent with scale invariance to at least $8\sigma$ \cite{Aghanim:2018eyx}; however, the exact correspondence between \eqref{nstow} and the scalar tilt depends on other slow-roll parameters as well.

\subsection{Lifetime of the universe when $\Lambda < 0$}

When $\Omega_{fld} + \Omega_{m}$ is greater than 1 the cosmological constant will be negative, and if $w_{0fld} > -1$ the universe will expand to a maximum size and then recontract.  Starting from a data set of approximately $200,000$ accepted points in the converged Monte Python chains that have a negative value for $\Omega_{\Lambda}$, we have solved the Friedmann equations numerically to estimate a lower bound on the lifetime of the universe.  We find no lifetimes less than about $9 H_{0}^{-1}$, and the average lifetime is approximately $29 H_{0}^{-1}$.  These correspond to 130 Gyr and 420 Gyr, respectively, in the Benchmark model with $H_0^{-1} = 14.4$ Gyr.  This result is the same whether the neutrino mass $\Sigma m_{\nu}$ is fixed to 0.06 eV or allowed to be more general.  In principle there is no upper bound on the lifetime, since the region $\Omega_{\Lambda} < 0$ is not excluded by the data.

\section{Conclusions and future directions}

To summarize the main results of this study, in accordance with the predictions of the simple harmonic universe class of models, we have investigated the experimental constraints on two negative energy sources, namely positive spatial curvature and negative c.c., and their interactions with exotic matter sectors.  For the curvature, it is known that the CMB Planck data alone mildly favor positive curvature at the $1\sigma$ level, but the error bars are still consistent with a flat universe at $2\sigma$, and the data become consistent with flatness at the $0.5$ per cent level after the CMB data are combined with BAO data.  It is also possible to positive curvature to affect the primordial spectrum and suppress the low multipoles even if $\Omega_k$ is otherwise unobservable today.  We find that the presence of matter with $w = -1/3$ partially removes the degeneracy with $H_0$ but does not greatly shift the constraints on curvature, while the presence of exotic matter with $w \approx -1$ tends to push the constraints toward zero or even positive $\Omega_k$.  If the cosmological constant is ultimately negative, modifications to $\Omega_{fld}, w_{0fld}$ in the dark energy sector may indicate that the lifetime of the current expanding phase of the Universe is bounded.

It may also be of interest to model competing effects on the primordial spectrum, such as those that may arise from a non-Bunch-Davies initial state or a potential with a rapidly running spectral tilt, and see how these affect the constraints on curvature and negative energy sources.  Modeling the effects on the bispectrum and higher-point functions could help us push further to disentangle these factors.  It would also be interesting to pursue further the constraints on this class of models coming from polarization, particularly if the model for the exotic matter sector allows for shear stresses. 

While the effects of exotic matter and $K_{eff} \ll 1$ affect the constraints on curvature at a level below the current size of the error bars, as upcoming experiments constrain the spatial curvature to the level of $10^{-3}$ and possibly further \cite{Leonard:2016evk, Takada:2015mma, Abazajian:2016yjj, DiValentino:2016foa}, it will be necessary to disentangle curvature from a host of similar and competing signals, which may include the modifications to the dark energy sector described here.
Regardless of the difficulty of searching for parameters of the primordial era, it is of great significance to detect or constrain these parameters as sensitively as possible, and in order to do so it will be necessary to consider all possible competing effects.  Theoretical models such as the SHU are useful in order to develop novel search templates, such as the reminder that $\Omega_k < 0$ is not forbidden at all, or that positive curvature may or may not be decoupled from the expansion history, or that the cosmological constant may ultimately be negative.  
Future experiments will continue to narrow the allowed parameter window and help us to constrain or test these possibilities.

\section*{Acknowledgments}

We thank Raphael Flauger, Colin Hill, Rostislav Konoplich, Constantine Theodosiou, and Claire Zukowski for very helpful discussions.  Part of this work was carried out using the resources of the Kakos Computer Cluster at Manhattan College, and we thank Kashifuddin Qazi, Stephen Romero, and Constantine Theodosiou for their support and assistance.  This work was supported in part by Office of the Provost and the Office of the Dean of the School of Science at Manhattan College, via the Jasper Summer Research Scholars program (PG) and by a Manhattan College Faculty Summer Grant (BH).  Preliminary versions of this work were presented at the Manhattan College 2018 Summer Research Symposium, the 4th International Conference on Particle Physics and Astrophysics in Moscow, Russia, the NJ American Association of Physics Teachers' annual meeting in November 2018, and the Society of Physics Students Zone 2 Regional Meeting in Rochester, NY, March 2019.  Preliminary results from this work also appear in the college research journal \textit{The Manhattan Scientist}.


\end{document}